\documentclass[12pt]{article}
\usepackage{epsfig}

\def\ni{\noindent}
\def\nl{\hfill\break}
\begin{document}
\begin{flushright}
hep-ph/0203021 \\
RAL-TR-2002-006 \\
29 Apr 2002 \\
\end{flushright}
\begin{center}
{\Large Neutrino Matter Effect Invariants and the Observables of}
\end{center}
\vspace{-7mm}
\begin{center}
{\Large Neutrino Oscillations}
\end{center}
\vspace{1mm}
\begin{center}
{P. F. Harrison\\
Physics Department, Queen Mary University of London\\
Mile End Rd. London E1 4NS. UK \footnotemark[1]}
\end{center}
\begin{center}
{and}
\end{center}
\begin{center}
{W. G. Scott\\
Rutherford Appleton Laboratory\\
Chilton, Didcot, Oxon OX11 0QX. UK \footnotemark[2]}
\end{center}
\vspace{1mm}
\begin{abstract}
\baselineskip 0.6cm
\noindent

We generalise our previous observation
on the invariance of the Jarlskog determinant
to matter effects in neutrino oscillations.
Within the context of standard neutrino
oscillation theory with matter effects,
we present the complete set of
(five) matter invariant observables
for neutrino propagation in matter.
We give some examples of their application.
\end{abstract}
\begin{center}
{\em To be published in Physics Letters B}
\end{center}

\footnotetext[1]{E-mail:p.f.harrison@qmul.ac.uk}
\footnotetext[2]{E-mail:w.g.scott@rl.ac.uk}

\newpage
\ni {\bf 1 Introduction}
\vspace{2mm}
\nl Neutrino oscillations violate lepton flavour conservation.
The amplitude $A_{\alpha\beta}$
for a neutrino of flavour $\alpha$
to be detected as a neutrino of flavour $\beta$
is given as a function of propagation distance $L$
by the (matrix) equation:
\begin{equation}
A = \exp(-iHL)
\label{expHam}
\end{equation}
where $H$ is the effective Hamiltonian in a flavour basis
(ie.\ in a weak basis in which the charged lepton mass matrices
are diagonal - this is defined up to an arbitrary phase
redefinition of the charged lepton fields).
We may take:
\begin{equation}
H=U{\rm diag}{(0,\Delta_{21},\Delta_{31})}U^{\dag}
\label{hamiltonian}
\end{equation}
where $U$ is the conventional MNS matrix
and the $\Delta_{ij}$ are defined by
\mbox {$\Delta_{ij}=(m^2_i-m^2_j)/2E$},
with $m_i$ the neutrino masses
and $E$ the neutrino energy.
In this paper, all calculations
are assumed to be in such a flavour basis.

We will find the following theorem useful.
Any function, $f(H)$, of a Hermitian
operator $H$, can be expanded using
Lagrange's formula \cite{shaw}:
\begin{equation}
f(H)=\sum_i f(\lambda_i) X^i
\label{theorem}
\end{equation}
where the $\lambda_i$ are the eigenvalues of $H$
and the $X^i$ are hermitian (projection)
operators given by:
\begin{equation}
X^i =
\frac{\prod_{j \neq i} (H-\lambda_j)}
                {\prod_{j \neq i} (\lambda_i-\lambda_j)}.
\label{lagrange}
\end{equation}
Since $H$ satisfies
its own characteristic equation,
the complete product
$\prod_i (H-\lambda_i)$
(with no factors dropped)
clearly vanishes.
It follows that the
$X^i$ are both left and right eigenvectors
of $H$ with eigenvalues $\lambda_i$,
ie.\ $(H-\lambda_i)X^i = X^i(H-\lambda_i) =0$.
With the normalisation defined as in Eq.~\ref{lagrange},
we have $(X^i)^2 = X^i$,
so the $X^i$ are indeed projection operators,
representing the (pure) mass-eigenstate
density-matrices.

The ancient formula Eq.~\ref{theorem} has been
used previously for the specific case of three-family neutrino
oscillations in eg.~Ref.~\cite{bargeretal}, and has
recently re-appeared in the same context in Ref.~\cite{ohlsen}.
Applied to Eq.~\ref{expHam} it gives:
\begin{equation}
A = \sum_i X^i \exp(-i\lambda_iL).
\label{amplitude}
\end{equation}
Any $X^i$ is readily computed
from the corresponding column
of the MNS matrix \cite{HPS33}:
\begin{equation}
X^i_{\alpha\beta} = U_{\alpha i}U^*_{\beta i}
\label{xi}
\end{equation}
(where no summation over $i$ is implied).
We note here that the elements of the matrices
$X^i$ are dependent on the phase-convention chosen
for the charged lepton fields
and are therefore not observables
(likewise, of course, neither are the
MNS matrix elements themselves).

Using Eq.~\ref{theorem}
we may compute arbitrary
powers of $H$ as follows:
\begin{equation}
H^r = \sum_i \lambda_i^r X^i.
\label{Hm}
\end{equation}
Given the eigenvalue spectrum,
Eq.~\ref{lagrange} determines
the matrices $X^i$ in terms of $H$,
whilst setting $r=1$ in Eq.~\ref{Hm}
yields the inverse relation, 
determining $H$ in terms of the $X^i$.
The case $r=0$ of Eq.~\ref{Hm}
expresses the unitarity of $U$.
For an $n\times n$ Hamiltonian,
there are only $n$ independent equations
of the form Eq.~\ref{Hm},
because the characteristic equation
can be used to write higher powers $r \ge n$
in terms of the $n$
lowest powers $H^r, 0 \le r < n$.

The squared amplitude $|A_{\alpha\beta}|^2$ for $\alpha \neq \beta$
($\alpha=\beta$) gives the appearance (survival)
probabilities as a function of propagation distance
(the dependence on neutrino energy is implicit 
in the factors $\Delta_{ij}$):
\begin{eqnarray}
P(\alpha \rightarrow \beta)=|A_{\alpha\beta}|^2 &=& \delta_{\alpha\beta}
-4\sum_{i<j}K^{ij}_{\alpha\beta}\sin^2{(\Delta_{ij}L/2)} \cr
&+&8J_{\alpha\beta}\sin{(\Delta_{12}L/2)}\sin{(\Delta_{23}L/2)}\sin{(\Delta_{31}L/2)}
\label{prob}
\end{eqnarray}
where
\begin{equation}
K^{ij}_{\alpha\beta}={\rm Re}(X^i_{\alpha\beta}X^{j*}_{\alpha\beta}),
\label{Kdef}
\end{equation}
and
\begin{equation}
J_{\alpha\beta}={\rm Im}(X^i_{\alpha\beta}X^{j*}_{\alpha\beta})
=\pm J~({\rm for}~\alpha \neq \beta) \quad {\rm or} \quad =0~({\rm for}~\alpha=\beta).
\label{Jdef}
\end{equation}
$J$ is Jarlskog's $CP$- and $T$-violating invariant \cite{jarls}.
Inspection of Eqs.~\ref{prob}-\ref{Jdef} indicates that the
phase-convention-independent products of the form
$X^i_{\alpha\beta}X^{j*}_{\alpha\beta}$ $(i\neq j)$ are
observables of the system \cite{bjorken}, the real parts
$K^{ij}_{\alpha\beta}$ parameterising $CP$-conserving oscillation amplitudes
and the imaginary parts $J_{{\alpha\beta}}$
parameterising $CP$ violation. We note that unitarity
imposes many constraints among the observables, 
$K^{ij}_{\alpha\beta}$ (and $J$),
and that once three of these are specified (such that at least
two different flavour indices are used) along with $J$,
the remaining ones may be determined via unitarity.

When neutrinos propagate in matter, the Hamiltonian in the
flavour basis is modified: $H \rightarrow H'$, with
\begin{equation}
H'=H+{\rm diag}(a,0,0)
\label{matterHam}
\end{equation}
where $a=\sqrt{2}G_F N_e$.
This modifies both the eigenvalues and eigenstates 
of the neutrino Hamiltonian
(in this paper, ``primed'' symbols will be used 
to denote the values of quantities when neutrinos propagate
in matter and unprimed symbols for their vacuum counterparts).

We recently \cite{hs2} pointed out that the determinant of the
commutator of the lepton mass matrices is invariant with respect to
matter effects,
\begin{equation}
{\rm Det}[D^2_{\ell},H']={\rm Det}[D^2_{\ell},H]
\label{commutator}
\end{equation}
where $D_{\ell}$ is the charged-lepton mass matrix,
which is diagonal by construction in this basis.
This allowed the derivation of a matter-invariant quantity,
expressible both as a function of elements of the 
effective Hamiltonian, and as a combination of the 
eigenvalue differences, $\Delta_{ij}$, and the $CP$ violation
parameter, $J$ \cite{hs2}:
\begin{eqnarray}
{\rm Im}(H'_{e\mu}H'_{\mu\tau}H'_{\tau e})
&=&{\rm Im}(H_{e\mu}H_{\mu\tau}H_{\tau e})\cr
=\Delta'_{12}\Delta'_{23}\Delta'_{31} J'
&=&\Delta_{12}\Delta_{23}\Delta_{31} J.
\label{triple}
\end{eqnarray}
This quantity has found application in several instances \cite{tphenom}.
It is not however, the only matter-invariant combination of
neutrino oscillation observables. Here we count the number of independent
such combinations and derive a complete set.

\vspace{7mm}
\ni {\bf 2 A Complete Set of Matter Invariant Observables}
\vspace{2mm}
\nl Clearly, in a flavour basis, 
the modification to the effective Hamiltonian 
induced by matter, Eq.~\ref{matterHam}, 
affects only the ``$ee$'' element:
\begin{equation}
H'_{ee}=H_{ee}+a
\label{hee}
\end{equation}
whereas the remaining eight elements
of $H$ are matter-invariant in this basis, ie.
\begin{equation}
H'_{\alpha\beta}=H_{\alpha\beta}, \quad \alpha\beta \ne ``ee".
\label{matterInv}
\end{equation}
The invariants of Eq.~\ref{matterInv} are not 
immediately useful, however, as the Hamiltonian 
(vacuum or matter) is not itself observable. 
It transforms
under (separate) phase redefinitions 
of the flavour eigenstates:
\begin{equation}
H\rightarrow \Phi H \Phi^{\dag}
\label{phasetr1}
\end{equation}
where
\begin{equation}
\Phi=\exp\{i~{\rm diag}(\phi_{e},\phi_{\mu},\phi_{\tau})\}
\label{phasetr2}
\end{equation}
so that
\begin{equation}
H_{\alpha\beta}\rightarrow e^{i(\phi_{\alpha}-\phi_{\beta})}H_{\alpha\beta}
\label{phasetr3}
\end{equation}
without any observable consequence.
The $X^i$ transform similarly.
If our invariants are 
to be readily applicable,
we must seek to relate 
observable quantities directly.

We can count the number of independent observables 
represented by $H$ as follows. 
An arbitrary $n\times n$ Hermitian matrix
comprises $n^2$ independent real parameters
($n$ on-diagonal, $n(n-1)$, off-diagonal).
Eqs.~\ref{phasetr1}-\ref{phasetr3} imply
that for the effective Hamiltonian,
$n-1$ of these are unobservable {\em relative} phases,
so that we would expect, at most,
$n^2-n+1$ observable quantities.
In addition however, subtraction of any multiple
of the identity from $H$ (a ``trace transformation'')
leaves all neutrino
oscillation observables invariant,
so that only eigenvalue {\em differences} are relevant,
leaving finally only \mbox{$n(n-1)$} independent observables 
in neutrino oscillations.
For the $3\times 3$ MNS oscillation phenomenology,
we have six observables, which are
conventionally considered to be three real angles,
one phase and two mass-squared differences. They could
equally well be taken to be three independent 
$K^{ij}_{\alpha\beta}$ (see Eq.~\ref{prob}), $J$ and 
two eigenvalue differences, $\Delta_{21}$ and $\Delta_{31}$, 
or alternatively six phase- 
and trace-invariant functions of elements of $H$ (see below).
When neutrinos propagate in matter, one parameter is modified,
Eq.~\ref{hee}, so that we may expect, in general, $n(n-1)-1$ matter
invariances (eg.~five, in the three-family case).

We seek relations analogous to 
Eqs.~\ref{hee}-\ref{matterInv} but involving only 
functions of the $H_{\alpha\beta}$ which are invariant under both
phase transformations and trace transformations.
Taking Eq.~\ref{triple} as a model,
we seek furthermore to write them in terms
of the observables of Eq.~\ref{prob}.
Six independent such relations should exist.
Relations which do not involve the ``ee'' flavour 
label will, in addition, be matter-invariant.
The off-diagonal elements,
$H_{\alpha\beta}$ $(\alpha \neq \beta)$,
are already invariant
under trace transformations, and we can form
phase-transformation invariants
simply by taking their moduli-squared:
\begin{eqnarray}
|H'_{\alpha\beta}|^2&=&|H_{\alpha\beta}|^2 \label{modSq1}\\
= |\sum_i \lambda'_i X^{'i}_{\alpha\beta}|^2 &=& |\sum_i \lambda_i X^{i}_{\alpha\beta}|^2 \label{modSq2}\\
= -\sum_{j<k} \Delta'^2_{jk}K'^{jk}_{\alpha\beta}&=&-\sum_{j<k} \Delta^2_{jk}K^{jk}_{\alpha\beta}, \quad \alpha \neq \beta
\label{modSq}
\end{eqnarray}
where we have used the relation,
Eq.~\ref{Hm} with $r=1$ to obtain Eq.~\ref{modSq2},
and with $r=0$ (unitarity) to obtain Eq.~\ref{modSq}.
The quantities in Eq.~\ref{modSq} 
are all observable (see eg.~Eq.~\ref{prob})
and the three independent combinations 
are each matter-invariant
(only three combinations because
$|H_{\alpha\beta}|=|H_{\beta\alpha}|$).
\footnote{The diagonal analogue of Eqs.~\ref{modSq1}-\ref{modSq}: 
$-\sum_{j<k} \Delta^2_{jk}K^{jk}_{\alpha\alpha}
=(H_{\alpha\alpha})^2-(H^2)_{\alpha\alpha}$
is not independent, the LHS equating through unitarity to:
$\sum_{j<k} \Delta^2_{jk}K^{jk}_{\alpha\beta}
+\sum_{j<k}\Delta^2_{jk}K^{jk}_{\alpha\gamma}, \alpha\ne\beta\ne\gamma$,
and the RHS being simply
$-|H_{\alpha\beta}|^2-|H_{\alpha\gamma}|^2$
}

In a complementary fashion,
the diagonal elements of $H$ are already phase-invariant
and we can form trace-transformation invariants by
taking differences, whereby:
\begin{eqnarray}
H'_{\mu\mu}-H'_{\tau\tau}
&=&H_{\mu\mu}-H_{\tau\tau} \cr
= (X^{'2}_{\mu\mu}-X^{'2}_{\tau\tau})\Delta'_{21}+(X^{'3}_{\mu\mu}-X^{'3}_{\tau\tau})\Delta'_{31}
&=&(X^2_{\mu\mu}-X^2_{\tau\tau})\Delta_{21}+(X^3_{\mu\mu}-X^3_{\tau\tau})\Delta_{31}\quad
\label{diag1}
\end{eqnarray}
which is matter-invariant, and
\begin{eqnarray}
H'_{ee}-Tr(H')/3
&=&H_{ee}-Tr(H)/3+2a/3 \cr
= (X^{'2}_{ee}-\frac{1}{3})\Delta'_{21}+(X^{'3}_{ee}-\frac{1}{3})\Delta'_{31}
&=&(X^2_{ee}-\frac{1}{3})\Delta_{21}+(X^3_{ee}-\frac{1}{3})\Delta_{31}+2a/3 \qquad
\label{diag2}
\end{eqnarray}
which may be said to be matter-covariant.

The three invariances of Eq.~\ref{modSq2} correspond 
to the ``sum-rules'' of Ref.~\cite{xing},
although they are derived in a different way here, 
without appealing to the commutator of the mass-matrices
\footnote {The quantities $Z_{ij}$ and $\tilde Z_{ij}$ defined in
Ref.~\cite{xing} are in fact simply the off-diagonal elements of the
effective Hamiltonian in vacuum and in matter respectively.}.
Their expression in terms of our observables, Eq.~\ref{modSq}, and the
relations of Eqs.~\ref{diag1} and \ref{diag2} are new. Our derivation
demonstrates that all the above invariances follow straightforwardly 
from the fact that the Wolfenstein term affects only the ``$ee$''
element of the effective Hamiltonian, Eqs.~\ref{hee}, \ref{matterInv}.

The five combinations
Eqs.~\ref{modSq}-\ref{diag2} all relate
$CP$-conserving observables
(the diagonal elements
$X^i_{\alpha\alpha}=|U_{\alpha i}|^2$ are observable).
The combination Eq.~\ref{triple}
parameterises $CP$ violation
and is therefore independent.
It is manifestly matter-, phase- and
trace-transformation invariant, as required.
We have therefore completed the derivation 
of six independent identities,
Eqs.~\ref{triple} and \ref{modSq}-\ref{diag2}, relating particular
combinations of matter-modified neutrino oscillation observables 
to their vacuum counterparts. 
Five of the combinations 
(Eqs.~\ref{triple}, \ref{modSq} and \ref{diag1})
are matter-invariant,
while one combination (Eq.~\ref{diag2}) 
depends in a simple way
on the matter-density. 
As regards neutrino oscillation phenomenology,
these six combinations constitute a complete set,
whereby any further matter-invariants 
(which can certainly be found,
eg.~${\rm Re}(H_{e\mu}H_{\mu\tau}H_{\tau e})$) 
must always be expressible in terms 
of those given here.

\vspace{7mm}
\ni {\bf 3 Some Applications}
\vspace{2mm}
\nl As a first application, we show briefly how the
formalism presented in the last Section 
applies to the simple case of
two neutrino families. 
In this case, there is only one independent
off-diagonal element of the effective Hamiltonian, $|H_{e\mu}|$,
which leads to the analogue 
of the matter-invariant in Eq.~\ref{modSq}:
\begin{eqnarray}
|H'_{e\mu}|^2&=&|H_{e\mu}|^2 \cr
= -\Delta'^2_{12}K'^{12}_{e\mu}&=&-\Delta^2_{12}K^{12}_{e\mu} \cr
\Rightarrow \Delta'^2_{12}\sin^2{2\theta'}&=&\Delta^2_{12}\sin^2{2\theta}
\label{modSq22}
\end{eqnarray}
while the two-family analogue of Eq.~\ref{diag2} is
\begin{eqnarray}
(X^{'2}_{ee}-\frac{1}{2})\Delta'_{21}&=&a/2+(X^2_{ee}-\frac{1}{2})\Delta_{21} \cr
\Rightarrow
(\sin^2\theta'-\frac{1}{2})\Delta'_{21}&=&a/2+(\sin^2\theta-\frac{1}{2})\Delta_{21} \cr
\Rightarrow -\cos{2\theta'}\Delta'_{21}&=&a-\cos{2\theta}\Delta_{21}
\label{diag22}
\end{eqnarray}
where we have used the definitions of the $X^i$ and
$K^{ij}_{\alpha\beta}$ in terms of the $2\times 2$ 
mixing matrix elements to
recover the standard results \cite{lang83}.
We have found one matter-invariant combination
of observables \cite{akhmedov}, and one which is matter-covariant,
the complete set for the two-family case 
(compared with five and one respectively 
for the three family case).

In the case of three families, 
we consider
the appearance probabilities
Eq.~\ref{prob} with $\alpha \neq \beta$,
in matter of arbitrary (uniform) density.
Expanding each term
in $\Delta'_{ij}L/2$ for $\Delta'_{ij}L/2$ small, we find:
\begin{eqnarray}
P'(\alpha \rightarrow \beta)&=&
-\sum_{i<j}K^{'ij}_{\alpha\beta}(\Delta'^2_{ij}L^2-\Delta'^4_{ij}L^4/24+...) \cr
&+&J'_{\alpha\beta}\Delta'_{12}\Delta'_{23}\Delta'_{31}L^3+...\cr
&=&|H'_{\alpha\beta}|^2L^2+{\rm Im}(H'_{e\mu}H'_{\mu\tau}H'_{\tau e})L^3+{\cal O}(\Delta'^4_{ij}L^4)+...\cr
&=&|H_{\alpha\beta}|^2L^2+{\rm Im}(H_{e\mu}H_{\mu\tau}H_{\tau e})L^3+{\cal O}(\Delta'^4_{ij}L^4)+...
\label{expansion}
\end{eqnarray}
We have used Eqs.~\ref{modSq} and \ref{triple}
to identify the coefficients respectively of 
the $CP$-conserving term of order $L^2$ and 
the $CP$-violating term of order $L^3$ with 
manifestly matter-invariant combinations of elements
of the effective Hamiltonian. Thus, they are determined
directly in terms of elements of the vacuum Hamiltonian.
The term of order $L^4$ is not
matter-invariant, 
and is the lowest-order term 
through which matter effects enter. 
The well-known \cite{bargeretal} result that
matter has no observable effect 
for small propagation distances,
is explained here by the fact 
that off-diagonal elements of $H$ 
are unmodified by matter.
The result, Eq.~\ref{expansion}, is valid
for all mass and mixing patterns, but is,
of course, useful only for small propagation distances.

A particular success of the
matter invariance, Eq.~\ref{triple}, 
is that it enables
the calculation of the universal $CP$ violating asymmetry
in matter of uniform density,
with no restriction to small $L$, 
knowing only the vacuum Hamiltonian and
the matter-modified eigenvalue differences \cite{hs2}
(which are readily calculated in terms of
vacuum quantities, \cite{bargeretal, xing2}):
\begin{equation}
P'(\alpha \rightarrow \beta)-P'(\beta \rightarrow \alpha)
={\rm Im}(H_{e\mu}H_{\mu\tau}H_{\tau e})
\frac{\sin{(\Delta'_{12}L/2)}\sin{(\Delta'_{23}L/2)}\sin{(\Delta'_{31}L/2)}}
{(\Delta'_{12}/2)(\Delta'_{23}/2)(\Delta'_{31}/2)}.
\label{triple3}
\end{equation}
This expression does not require knowledge 
of the matter-modified eigenstates (ie.~mixing parameters), 
whose dependence is
contained completely within the matter-invariant factor
${\rm Im}(H_{e\mu}H_{\mu\tau}H_{\tau e})$.
A comparison of the $CP$-conserving and $CP$-violating  
terms in Eq.~\ref{expansion} 
leads us to suspect that, perhaps more generally, 
the matter-invariant quantities 
$|H_{\alpha\beta}|^2$ play a similar role
for $CP$-conserving observables 
to that played by ${\rm Im}(H_{e\mu}H_{\mu\tau}H_{\tau e})$
for $CP$-violating observables.
It is interesting therefore to enquire whether 
a simplification similar to that of Eq.~\ref{triple3}
exists also for the $CP$-conserving parts of 
$P(\alpha \rightarrow \beta)$,
in terms of the matter-invariants $|H_{\alpha\beta}|^2$.

Some guidance is obtained by 
returning briefly to the two-family case.
Consider Eq.~\ref{prob} for the 
$e \rightarrow \mu$ appearance probability
in matter of uniform density:
\begin{eqnarray}
P'(e \rightarrow \mu)=|A'_{e\mu}|^2 &=&
  -4K^{'12}_{e\mu}\sin^2{(\Delta'_{21}L/2)} \cr
&=&|H_{e\mu}|^2\frac{\sin^2{(\Delta'_{21}L/2)}}{(\Delta'_{21}/2)^2}
\label{newprob22}
\end{eqnarray}
where we have used Eq.~\ref{modSq22} to write the probability as the
product of a matter-invariant part $(|H'_{e\mu}|^2=|H_{e\mu}|^2)$ and a part 
with a generic dependence on eigenvalue differences of the form 
$[\sin{(\Delta' L/2)}/(\Delta'/2)]^n$.
This exact expression, Eq.~\ref{newprob22}, is clearly analogous 
to Eq.~\ref{triple3} in the sense we require.

We do not a priori expect the three-family case 
to be as straightforward,
but it will turn out
that much of the above simplicity is retained,
especially 
in the case of a hierarchical neutrino spectrum
(as seems to be preferred experimentally).
We may start again with Eq.~\ref{prob},
and assume a hierarchical spectrum,
$\Delta'_{21} << \Delta'_{31}$, 
$\Delta'_{31}\simeq\Delta'_{32}\equiv\Delta'$ 
for all matter-densities considered. 
Then in the limit
$\Delta'_{21}L/2 << 1$ (which is less restrictive than
the limit in which Eq.~\ref{expansion} applies),
we find for the leading oscillation of the
appearance probability (disappearance probabilities
are anyway readily obtained from appearance
probabilities through unitarity):
\begin{eqnarray}
P'(\alpha \rightarrow \beta)=|A'_{\alpha\beta}|^2 &\simeq& 
-4(K^{'31}_{\alpha\beta}+K^{'32}_{\alpha\beta})\sin^2{(\Delta'L/2)} \cr
&=&4|X^{'3}_{\alpha\beta}|^2\sin^2{(\Delta'L/2)}
\label{approx1}
\end{eqnarray}
where the last step follows from unitarity. We also require that
\mbox{$|X^{'2}_{\alpha\beta}|\Delta'_{21} << |X^{'3}_{\alpha\beta}|\Delta'_{31}$}
for all matter-densities considered, in order that the leading
oscillation dominates. This will generally be true, 
for the hierarchical spectrum we have assumed,
but may not be in some special cases eg.~if the mixing matrix elements 
also have an extreme hierarchy, see below.

Applying the same approximations to
Eq.~\ref{modSq} we have:
\begin{eqnarray}
|X^{'3}_{\alpha\beta}|^2 \Delta'^2 
=-(K^{'31}_{\alpha\beta}+K^{'32}_{\alpha\beta})\Delta'^2
\simeq |H'_{\alpha\beta}|^2
=|H_{\alpha\beta}|^2 
\simeq -(K^{31}_{\alpha\beta}+K^{32}_{\alpha\beta})\Delta^2
=|X^{3}_{\alpha\beta}|^2\Delta^2
\label{simpInv}
\end{eqnarray}
(ie.~the combination $|X^{'3}_{\alpha\beta}|\Delta'$
is a matter-invariant,  as long as
$|X^{'2}_{\alpha\beta}|\Delta'_{21} << |X^{'3}_{\alpha\beta}|\Delta'_{31}$
remains valid, as already assumed).
Combining Eqs.~\ref{approx1} and \ref{simpInv}, we find that
\begin{equation}
P'(\alpha \rightarrow \beta)=|A'_{\alpha\beta}|^2 \simeq
|H_{\alpha\beta}|^2\frac{\sin^2{(\Delta' L/2)}}{(\Delta'/2)^2}
\label{approx}
\end{equation}
which generalises 
Eq.~\ref{newprob22}
to the three-family case, when the
specified approximations apply.
\footnote{
The (possibly realised \cite{HPS5}) case $|U_{e3}|^2=0$, is a specific
case which violates the assumption
$|X^{'2}_{\alpha\beta}|\Delta'_{21} << |X^{'3}_{\alpha\beta}|\Delta'_{31}$
for $\alpha=e$ or $\beta=e$, because then, $X^{'3}_{\alpha\beta}=0$,
and the leading oscillation vanishes. However, in this case, the
long-wavelength oscillation becomes dominant, and Eqs.~\ref{approx1}
and \ref{approx} become exactly true, with the
substitutions $|X^{'3}_{\alpha\beta}| \rightarrow |X^{'2}_{\alpha\beta}|$
and $\Delta'\equiv\Delta'_{21}$ etc.}

Thus Eq.~\ref{approx} comes close
(as close as we believe is possible) 
to being the analogue 
of Eq.~\ref{triple3} 
for the CP-conserving case.
Where valid,
Eq.~\ref{approx}
retains all the   
calculational expediency
previously recognised
in connection with
Eq.~\ref{triple3},
for the CP-violating case.
We should say that in general
(for arbitrary mass-squared differences, 
mixing angles and
propagation lengths)
no such analogy exists 
between the $CP$-conserving 
and the $CP$-violating
contributions.
The simplification
exemplified by Eq.~\ref{triple3} 
is a consequence of the universality 
(for three families only)
of the CP-violating term,
as embodied in the
Jarlskog parameter, $J$ \cite{jarls}, 
but in general, no such universality exists 
for the CP-conserving $K^{ij}_{\alpha\beta}$ coefficients
(the two-family case being an exception). However, for 
phenomenologically realistic oscillation parameters 
(see eg.~\cite{HPS5}), Eq.~\ref{approx} is certainly 
applicable to the leading oscillation of 
$P'(\mu \rightarrow \tau)$ (and of $P'(\tau \rightarrow \mu)$) 
for most energies of interest, and, unless it turns out that 
$|X^{3}_{e\beta}|\Delta_{31} \sim |X^{2}_{e\beta}|\Delta_{21}$
for $\beta=\mu, \tau$ 
\mbox {($|X^{3}_{e\beta}|$ depends on the presently
poorly determined $U_{e3}$)}, it should also be relevant for 
appearance probabilities involving electron neutrinos.

\vspace{7mm}
\ni {\bf 4 Relation to Standard Formalism}
\vspace{2mm}
\nl In order to relate our matter-covariant formulation
to the standard one in terms of mixing angles, we first need to remove 
the implicit dependence of our matter-invariants 
on the neutrino energy (which has been left in until now 
in the interests of a concise discussion).
We note that all the preceding equations
in this paper remain unaltered, if we make the energy
re-scaling $H \rightarrow 2EH\equiv {\cal H}$,
$\Delta_{ij} \rightarrow 2E\Delta_{ij}\equiv \Delta m^2_{ij}$,
$\lambda_i \rightarrow 2E\lambda_i$,
$a \rightarrow 2Ea\equiv {\cal A}$, $L \rightarrow L/2E$ and likewise
for all primed quantities. The observables $K^{ij}_{\alpha\beta}$,
$J$, and their matter-modified counterparts remain unchanged throughout,
as do the oscillation probabililties.

We can now give the relationships between our energy-scaled 
matter-invariants (and covariant) (Eqs.~\ref{triple} and 
\ref{modSq1}-\ref{diag2}) and the six 
parameters conventionally used to describe
neutrino oscillation phenomenology,
$\theta_{12}$, $\theta_{23}$, $\theta_{13}$, $\delta$,
$\Delta m^2_{21}$ and $\Delta m^2_{31}$.
Defining $D=\Delta m^2_{31}-\Delta m^2_{21}s^2_{12}$
and $F=\frac{1}{8}S_{12}S_{23}S_{13}c_{13}$, we have:
\begin{equation}
{\cal H}_1\equiv|{\cal H}_{e\mu}|^2=c^2_{13}[s^2_{23}s^2_{13}D^2
+\frac{1}{4}c^2_{23}S^2_{12}(\Delta m^2_{21})^2]+2FD\Delta m^2_{21}c_{\delta}
\label{hemu}
\end{equation}
\begin{equation}
{\cal H}_2\equiv|{\cal H}_{e\tau}|^2=c^2_{13}[c^2_{23}s^2_{13}D^2
+\frac{1}{4}s^2_{23}S^2_{12}(\Delta m^2_{21})^2]-2FD\Delta m^2_{21}c_{\delta}
\label{hetau}
\end{equation}
\begin{equation}
{\cal H}_3\equiv|{\cal H}_{\mu\tau}|^2=\frac{1}{4}\{[c^2_{13}S_{23}D
-(C_{12}S_{23}+S_{12}C_{23}s_{13}c_{\delta})\Delta m^2_{21}]^2
+S^2_{12}s^2_{13}s^2_{\delta}(\Delta m^2_{21})^2\}
\label{hmutau}
\end{equation}
\begin{equation}
{\cal H}_4\equiv{\rm Im}({\cal H}_{e\mu}{\cal H}_{\mu\tau}{\cal H}_{\tau e})
=\Delta m^2_{21}\Delta m^2_{31}(\Delta m^2_{31}-\Delta m^2_{21})Fs_{\delta}
\label{hemutau}
\end{equation}
\begin{equation}
{\cal H}_5\equiv{\cal H}_{\mu\mu}-{\cal H}_{\tau\tau}=C_{23}[C_{12}\Delta m^2_{21}-c^2_{13}D]
-S_{12}S_{23}s_{13}c_{\delta}\Delta m^2_{21}
\label{hmumutautau}
\end{equation}
\begin{equation}
{\cal H}_0\equiv
{\cal H}_{ee}-\frac{1}{3}Tr{\cal H}=D(s^2_{13}-\frac{1}{3})
-\frac{1}{3}C_{12}\Delta m^2_{21}
\label{heetr}
\end{equation}
%
%
where we have used the conventional parameterisation \cite{PDG} and
$c_{ij}=\cos{\theta_{ij}}$, \mbox{$s_{ij}=\sin{\theta_{ij}}$},
$C_{ij}=\cos{2\theta_{ij}}$, $S_{ij}=\sin{2\theta_{ij}}$ etc. 
Eqs.~\ref{hemu}-\ref{hmumutautau}
remain true in matter (with primed quantities on the RHS), 
as well as in vacuum and 
show that (sums of) products of 
(matter-density-dependent) oscillation amplitudes and 
mass-squared-differences are matter-invariant, 
thereby accounting
for the folklore ``no-win'' theorem \cite{banuls},
that increased mixing-angles in matter seem 
to be compensated by reduced masses and
hence longer oscillation lengths, and vice-versa.
The covariant quantity on the LHS of Eq.~\ref{heetr} transforms 
to the matter case as
${\cal H}'_0\equiv{\cal H}'_{ee}-Tr({\cal H'})/3={\cal H}_0+2{\cal A}/3$,
where \mbox {${\cal A}=2\sqrt{2}G_FN_eE$}, while the quantities on the
RHS take their density-dependent values.

The usual six parameters
($\theta_{12}, \theta_{23}, \theta_{13}, \delta,
\Delta m^2_{21}~{\rm and}~\Delta m^2_{31}$),
are popular for their intuitive relevance,
but they {\em all}~have complicated dependencies
on the density of the matter through which the neutrinos
propagate \cite{bargeretal} \cite{zaglauer}.
Different experiments operating over different baselines
at different energies are all subject to different modifications
of each of these conventional parameters due to matter effects. 
However, in each case, only one additional parameter is actually present,
the (path-averaged) Wolfenstein term, ${\cal A}$. 
By contrast, as we have seen, 
the five combinations of observables,
Eqs.~\ref{hemu}-\ref{hmumutautau}
are all matter-invariant, while the sixth observable,
Eq.~\ref{heetr},
has a very simple dependence on the matter density.
Clearly, these quantities could be used as the basis for a new parameterisation
of neutrino oscillation phenomenology. This is especially true for
the matter-invariants involving only off-diagonal elements:
$|{\cal H}_{\alpha\beta}|^2 (\alpha \neq \beta)$ and 
${\rm Im}({\cal H}_{e\mu}{\cal H}_{\mu\tau}{\cal H}_{\tau e})$, which
are determined directly in appearance experiments as the 
amplitudes of the leading $CP$-even oscillations (given a strong-enough
mass-hierarchy) and the
$CP$-asymmetry respectively. Not only are they matter-invariant, but they
are more directly observed even than the conventional mixing angles.

To some degree then, one may say
that the problem of deconvoluting
matter effects on the standard mixing parameter set
is a problem of man's own making, which is solved 
(at least in the uniform-density case we have considered)
by simply switching to a different parameter set.
While it is probably unrealistic
to assume that the new parameterisation
proposed here will be immediately adopted universally,
we believe that our parameters may have a practical value,
beyond their evident conceptual importance
as matter-invariant observables.

\vspace{7mm}
\ni {\bf Acknowledgements}
\vspace{2mm}
\nl It is a pleasure to thank R. Edgecock for useful comments. This work
was supported by the UK Particle Physics and Astronomy Research Council
(PPARC).

\end{document}